\documentclass[twocolumn,final]{svjour3}
\smartqed  
\usepackage{graphicx}
\begin{document}

\title{Systematic studies of rubidium-exposed surfaces by X-ray photoelectron spectroscopy and light-induced atom desorption}
\author{Atsushi Hatakeyama \and
Sae Watashima \and
Ayumi Yamaguchi \and
Tomohiro Ikeno \and
Hiroyuki Usui \and
Ko Kato \and
Ken-ichi Harada \and
Yasuhiro Sakemi \and
Emilio Mariotti }

\institute{Atsushi Hatakeyama \at Department of Applied Physics, Tokyo University of Agriculture and Technology, Tokyo 184-8588, Japan 
\and
Sae Watashima \at
Department of Applied Physics, Tokyo University of Agriculture and Technology, Tokyo 184-8588, Japan
\and
Ayumi Yamaguchi \at
Department of Applied Physics, Tokyo University of Agriculture and Technology, Tokyo 184-8588, Japan
\and
Tomohiro Ikeno \at
Department of Applied Physics, Tokyo University of Agriculture and Technology, Tokyo 184-8588, Japan
\and
Hiroyuki Usui \at
Department of Organic and Polymer Materials Chemistry, Tokyo University of Agriculture and Technology, Koganei, Tokyo 184-8588, Japan
\and
Ko Kato \at
Cyclotron and Radioisotope Center (CYRIC), Tohoku University, 6-3 Aramaki-aza Aoba, Aoba-ku, Miyagi 980-8578, Japan
\and
Ken-ichi Harada \at
Cyclotron and Radioisotope Center (CYRIC), Tohoku University, 6-3 Aramaki-aza Aoba, Aoba-ku, Miyagi 980-8578, Japan
\and
Yasuhiro Sakemi \at
Center for Nuclear Study, The University of Tokyo, Wako, Saitama 351-0198, Japan
\and
Emilio Mariotti \at
Department of Earth, Environmental and Physical Sciences, University of Siena, Italy
}

\date{Received: date / Accepted: date}

\maketitle

\begin{abstract}
We systematically investigated various types of surfaces on which rubidium (Rb) atoms were deposited by X-ray photoelectron spectroscopy (XPS) and measured the light-induced atom desorption (LIAD) from those surfaces. The main surfaces of interest included synthetic quartz, yttrium metal, and paraffin. The Rb atoms deposited on quartz and yttrium surfaces by exposure to Rb vapor at room temperature were detected by XPS. Quartz is originally silicon dioxide. The yttrium surfaces were also oxidized, and Rb atoms reacted with oxygen on both surfaces. Conversely, Rb deposition was observed only at low temperatures on paraffin. Specifically, Rb atoms deposited on paraffin, which is not an oxygen compound, also formed oxygen compounds under ultrahigh vacuum conditions by reaction with the background gas. All examined surfaces showed a similar light wavelength or photon energy dependence, such that the LIAD rates increased with decreasing light wavelength. We presume that some types of compounds of alkali metal and oxygen can be ubiquitous sources for LIAD from many types of surfaces of alkali-metal vapor cells.
\end{abstract}

\section{Introduction}
Upon light illumination, alkali-metal atoms are ejected from the surface on which the atoms are deposited. This photo-stimulated desorption, referred to as light-induced atom desorption (LIAD), has been observed in various types of alkali-metal vapor cells. Alkali-metal vapor cells are widely used in atomic spectroscopy experiments, including laser cooling and trapping; LIAD is particularly useful for quickly loading cells with a sample atomic gas on demand~\cite{And01,Han01}.

\begin{figure}
\includegraphics[width=8cm]{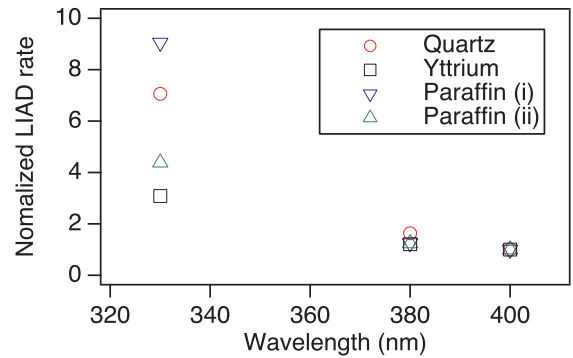}
\caption{(Color online) Wavelength dependence of the light-induced atom desorption (LIAD) rate of Rb atoms from several substrates. The LIAD rates are normalized by the photon fluxes and the rates at 400~nm. Paraffin (i) and (ii) are specified in Fig.~\ref{time-lapse}. The quartz data are the same as in Ref.~\cite{Kum16}.}
\label{summary}
\end{figure}

LIAD was first observed in alkali-metal vapor cells coated with polydimethylsiloxane (PDMS) in the early 1990s~\cite{Goz93,Meu94}. Subsequently, LIAD has been observed in various types of cells, including those coated with liquid helium~\cite{Hat00,Hat02}, octamethylcyclotetrasiloxane~\cite{Mar01}, paraffin~\cite{Ale02}, and octadecyltrichlorosilane~\cite{Cap07}, and uncoated cells made of glass and metal~\cite{And01,Han01}. LIAD experiments using alkali-metal vapor cells have focused on the desorption yield as a function of the light intensity and wavelength, as well as the loading dynamics of vapor density following LIAD~\cite{Goz93,Meu94,Ale02,Atu99,Kle06,Krz09_1,Krz09_2}. Except for liquid-helium-coated surfaces, the observed LIAD has similar light-wavelength or photon-energy dependence, such that LIAD yields monotonically increase with decreasing wavelength. A summary of our results regarding the wavelength dependencies of various surfaces is shown in Fig.~\ref{summary}.
However, detailed microscopic mechanisms have not been studied, partly because such studies require surface characterization, which is difficult for sealed cells. Accordingly, to study LIAD properties it is useful to place sample substrates in a vacuum chamber, in combination with surface analysis techniques. Such studies, however, have been limited to date~\cite{Bre04,Kit12,Hib13}. 

Detailed microscopic LIAD mechanisms should depend on surface materials and conditions. Our past X-ray photoelectron spectroscopy (XPS) study \cite{Kum16} involving two types of glass, synthetic quartz and Pyrex (a commercial borosilicate glass), revealed different states and LIAD properties of the two Rb-deposited surfaces. As another example of material dependence, time-of-flight measurements for desorbed atoms from PDMS films have indicated that the desorption process includes a diffusion component~\cite{Bre04,Kas04}. 

Here, we examined LIAD to identify commonalities among various types of surfaces. The investigated surfaces included synthetic quartz, yttrium metal, and paraffin. We presume that some types of oxygen compounds can be a common source for LIAD on the basis of the following observations. The main component of glass is silicon dioxide (SiO$_2$), together with other ingredients (e.g., B$_2$O$_3$, Al$_2$O$_3$, and Na$_2$O for Pyrex). Many metal surfaces, including the yttrium investigated in this research, are often oxidized. XPS measurements revealed that Rb atoms were easily deposited at room temperature, indicating a reaction between oxygen and Rb. The investigated polymer films, paraffin, and PDMS did not adsorb Rb atoms at room temperature; Rb atoms on the surface were deposited and detectable by XPS only at low temperature. Deposited Rb formed compounds containing oxygen. Paraffin is specifically an oxygen-free material; thus, the oxygen bound to Rb in this context originated from the residual background gas.
LIAD observed from various surfaces exhibited qualitatively similar properties at room temperature: a linear light-power dependence and increasing yield with increasing photon energy (or decreasing light wavelength), as shown in Fig.~\ref{summary}.

This paper is organized as follows. In Section II, we first explain the experimental apparatus. In Section III, we discuss the general features of deposited Rb as a function of surface temperature for quartz substrates. LIAD from Rb aggregates that were formed and stable only at low temperature is shown, together with LIAD at room temperature and above. LIAD from Rb-doped quartz is also presented. In Section IV, we present our experimental results for yttrium metal, which is particularly important in experiments involving radioactive francium. In Section V, we describe our investigation of polymer films, including detailed analysis of LIAD from Rb-oxygen compounds on oxygen-free paraffin film. The paper is concluded in Section VI.

\section{Experimental apparatus}
\begin{figure}
\includegraphics[width=8cm]{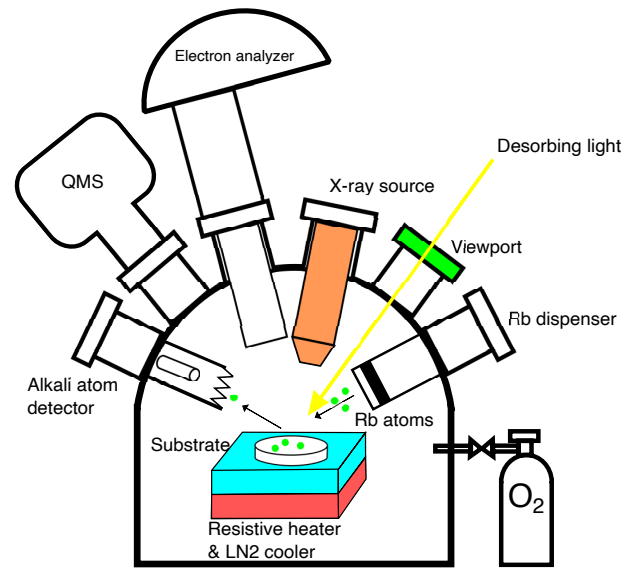}
\caption{(Color online) Schematic drawing of the experimental setup. LN2: liquid nitrogen; QMS: quadrupole mass spectrometer; O$_2$: oxygen gas.}
\label{apparatus}
\end{figure}

Figure~\ref{apparatus} shows a schematic diagram of the main experimental apparatus used throughout this study.
A detailed description of the apparatus was provided in Ref.~\cite{Kum16}.
Here, we provide a brief explanation of the set-up. The sample substrate on which Rb atoms were to be deposited was mounted on the sample stage, the temperature of which was controlled using liquid nitrogen (LN2) and a resistive heater from 130 to 670~K. Rb atoms were deposited onto the substrate from an Rb dispenser (SAES Getters S.p.A., Milan, Italy). The substrate surface, typically a few nanometers in depth, was analyzed by XPS using an X-ray source (Al K$\alpha$ ray) and an electron analyzer. Note that XPS does not detect hydrogen. LIAD light from a color-filtered xenon (Xe) lamp (wavelength range: 330 to 700~nm) or from an ultraviolet laser diode (wavelength: 375~nm) was incident on the substrate to induce Rb atom desorption. Typical light powers were 10 and 100~mW, and typical irradiation areas were 3 and 1~cm$^2$ for the Xe lamp and ultraviolet laser diode, respectively. The desorbed Rb atoms were detected sequentially with a homemade Langmuir-Taylor detector, which was equipped with a platinum filament to ionize neutral Rb atoms and a secondary electron multiplier. The pressure and components of the background gas in the vacuum chamber were monitored using a Bayard-Alpert ionization gauge and a quadrupole mass spectrometer. The base pressure was $1\times10^{-7}$~Pa. The typical main background gas components were CO$_2$ (27\%), N$_2$ (26\%), H$_2$ (14\%), CO (13\%), H$_2$O (8.2\%), and O$_2$ (6.3\%).
Oxygen gas was also introduced into the chamber to oxidize the Rb atoms.

\section{General features of the deposited Rb and LIAD: quartz substrates, as examples}
The amount of Rb atoms deposited on a surface after exposure to Rb vapor depends on the substrate material. Among the materials that we have studied, glass (synthetic quartz and Pyrex) and metal (yttrium) adsorbed Rb atoms, which were detected by our XPS system (on the order of $10^{12}$-$10^{13}$~cm$^{-2}$ surface density or 10$^{-1}$ - 10$^{-2}$ surface concentration~\cite{Kum16}) after exposure to Rb vapor at room temperature, whereas no Rb was observed on the polymer (paraffin and PDMS) surfaces.
Rb aggregates or films that form on a substrate are unstable and tend to desorb at room temperature under ultrahigh vacuum conditions, due to the high vapor pressure of Rb ($10^{-5}$~Pa).
Exposure to Rb vapor at low temperatures was thus required to obtain visibly thick Rb films on substrates for moderate rates of Rb impingement in our experiments.

\begin{figure}
\includegraphics[width=8cm]{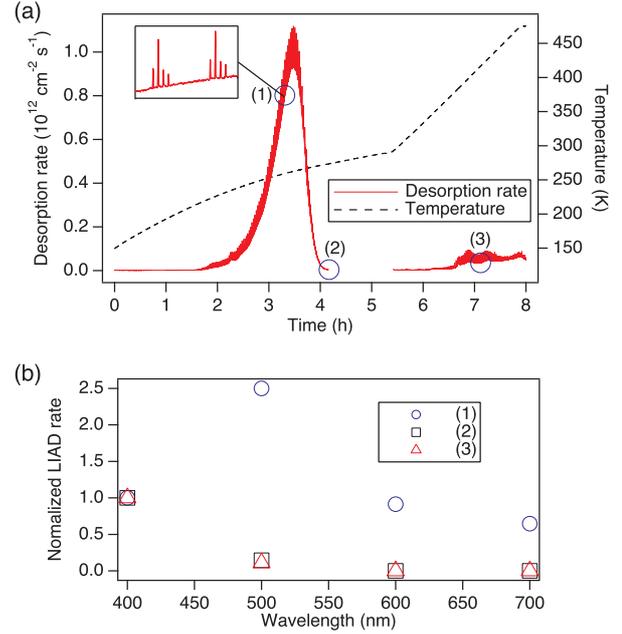}
\caption{(Color online) (a) The thermally and light-induced desorption signals recorded as the quartz substrate temperature increased. The baseline that peaks at 3.5~h (260~K) and increases after 6~h (330~K) corresponds to the thermal desorption of Rb. Spikes on the baseline correspond to the light-induced desorption of Rb by irradiation of one set of four Xe-lamp pulses with wavelengths of 400, 500, 600, and 700~nm at 1-s duration and 8-s intervals every 100~s (see magnified signal in the inset). Photon flux was identical ($6 \times 10^{15}$~s$^{-1}$) for all pulses, corresponding to 3~mW at 400~nm. The numbers (1), (2), and (3) represent the points at which the wavelength dependencies shown in (b) were measured. (b) Wavelength dependencies of the LIAD rate measured at points (1), (2) and (3), as indicated in (a).}
\label{quartz}
\end{figure}

Figure~\ref{quartz}(a) shows temporal changes in the thermally and light-induced desorption of Rb with increasing temperature after the formation of a metal Rb film by Rb deposition at 130~K. Thick Rb film formation was also confirmed by the Rb 3d XPS peak, which exhibited a high binding energy tail originating from the plasmon excitation typical of metal films, as shown in Fig.~4 of Ref.~\cite{Kum16}. Rb atoms began desorbing thermally at 1.5~h (210~K) and a large peak in thermal desorption was observed at 3.5~h (260~K). The Rb desorption rates were calibrated with the rates from a thick Rb film, as described in Ref.~\cite{Kit12}. 
Thermal desorption decreased before the temperature reached room temperature. This indicated complete desorption of the Rb film. Depletion of the Rb film was also confirmed by the disappearance of the plasmon loss peaks in the Rb 3d XPS peak, as shown in Fig.~4 of Ref.~\cite{Kum16}.
Further elevation of the temperature above room temperature led to additional enhancement of thermal desorption.

LIAD is evident in Fig. \ref{quartz}(a) as spikes on the thermal desorption curves due to the pulsed application of the Xe lamp. The LIAD rate was high when the thermal desorption was active at 260~K and above room temperature; however, the wavelength-dependence differed for metallic Rb (the Rb film) at low temperatures and non-metallic Rb at room temperature and above. LIAD exhibited a resonant-like dependence on wavelength for the Rb film, as indicated by the LIAD rate (1) in Fig.~\ref{quartz}(b); it showed a monotonic increase with decreasing light wavelength, as indicated by the LIAD rates (2) and (3) in Fig.~\ref{quartz}(b). This resonance behavior in the visible range has been observed for alkali-metal films and clusters~\cite{Hoh88}. Our main focus in this research was LIAD from non-metallic-Rb surfaces at room temperature and above, in which the LIAD yield increased monotonically with decreasing light wavelength, with a linear dependence on light power. For such surfaces, our XPS measurements revealed that Rb deposition changed the shape of the O 1s peak, thus indicating the reaction of Rb with the surface O~\cite{Kum16}.

The microscopic mechanisms of LIAD are beyond the scope of this paper. However, it is instructive to refer to several past studies in this context. LIAD of alkali-metal atoms from silicate surfaces has been studied extensively by Maedy and his coworkers in an attempt to identify the source of alkali-metal atoms contained in the atmospheres of the Moon and the planet Mercury~\cite{Yak99,Yak00,Yak03,Dom04}.
Their experimental and theoretical studies suggest that LIAD near a specific threshold (4~eV for sodium) likely originates from electron transfer from the non-bridging oxygen at the SiO$_2$ surface to cationic alkali-metal adsorbates. 

Of particular interest is that our results showed that quartz substrates with extensive exposure to Rb remained ``LIAD-active'', despite exposure to air and rinsing with water. We reacted quartz substrates with Rb metal in evacuated glass tubes at approximately 600~K for 1-2 h. We then rinsed the substrates to remove surface Rb compounds, which were brown in color. We used XPS to confirm that Rb remained on the surface after the rinse. Desorption of Rb atoms from the quartz substrate under light illumination was observed at approximately 500~K, at which LIAD was especially active, as also shown in Fig.~\ref{quartz}(a). Thus far, we have not been able to obtain sufficient LIAD yields at room temperature for laser cooling or atomic spectroscopy applications using such Rb-doped quartz substrates; however, it may be feasible to use Rb-doped, LIAD-active materials as a new type of atom source in the future.

\section{Yttrium metal}

\begin{figure}
\includegraphics[width=8cm]{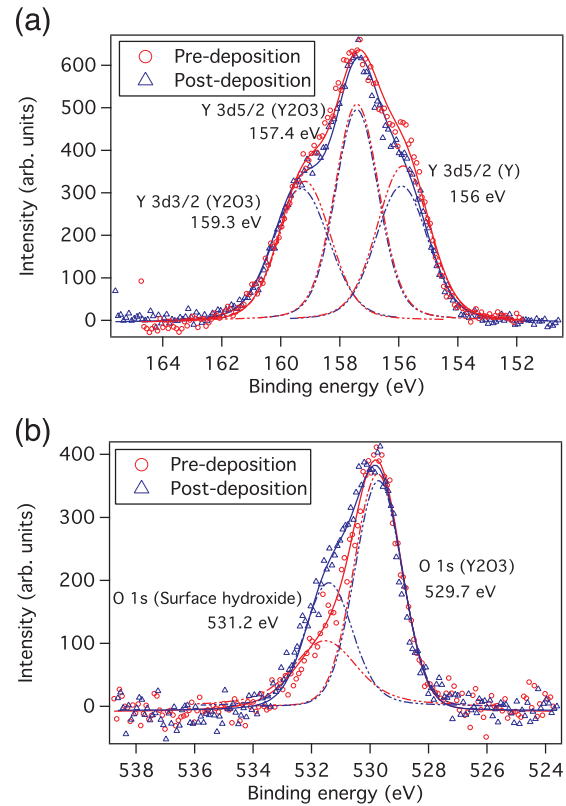}
\caption{(Color online) X-ray photoelectron spectroscopy (XPS) peaks for (a) Y 3d and (b) O 1s of the Y surface before and after Rb deposition. Baselines have been subtracted from the original data for the fitting. The peak assignment follows Ref.~\cite{Bar01}, except for the Y 3d$_{5/2}$ (Y) peak at 156~eV, which is taken from the literature~\cite{Bri03}.}
\label{Y_xps}
\end{figure}

\begin{figure}
\includegraphics[width=8cm]{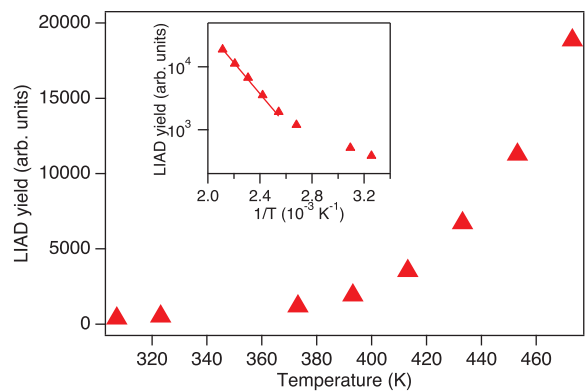}
\caption{(Color online) Temperature dependence of the LIAD yield from Rb-deposited Y metal for light with a wavelength of 375~nm. Inset shows an Arrhenius plot.}
\label{Ytemp_dep}
\end{figure}

Yttrium (Y) is a metal with a low work function of 3~eV. It is thus used to neutralize impinging ions, such as in the neutralization of francium (Fr) ions. Fr is the heaviest known alkali metal. All Fr isotopes are unstable and do not exist in nature. However, their comparatively long lifetimes (i.e., minutes) and heavy alkali metal properties offer unique spectroscopy opportunities involving neutral Fr atoms for fundamental physics tests, such as precision measurements of parity non-conservation effects~\cite{Sim96,Sim98,Bou08,Mar14,Aok17} and the search for permanent electric dipole moments~\cite{Byr99,Kaw16}. Y metal has been used in such experiments as a neutralizer of Fr ions originally produced by a nuclear fusion reaction and extracted as ions. The release of Fr as neutral atoms from Y is usually performed by heating the Y metal (e.g., to 970~K)~\cite{Sim96}. It would be beneficial to use LIAD to enhance the neutral Fr yield at lower temperatures to lower the background pressure. Accordingly, LIAD from Rb- or Fr-ion-implanted Y metal has been reported~\cite{Cop14}. LIAD from PDMS-coated surfaces has also been used to load a magneto-optical trap (MOT) of Fr atoms~\cite{Agu17}. 

In this study, we examined Y substrates exposed to Rb vapor at room temperature by XPS and LIAD. The chemical properties of Rb and Fr atoms are similar, and our knowledge regarding Rb acquired in this study can be applied to Fr. XPS spectra taken for a Y metal plate after heating at 670~K for 46 h are shown in Fig.~\ref{Y_xps}. The binding energy was calibrated by adjustment of the Y 3d$_{5/2}$ peak to 157.4~eV for Y$_2$O$_3$~\cite{Bar01}. There is oxygen on the Y surface, mainly as Y$_2$O$_3$, with a smaller high-energy peak assigned to the surface hydroxide, the peak of which was much larger before heating. Rb was detected by XPS after Rb exposure at room temperature. Post-deposition spectra for the O 1s peak in Fig.~\ref{Y_xps}(b) show that Rb deposition increased O concentrations on the surface, indicating that Rb also formed oxygen compounds by reaction with the background gas (probably O$_2$ and/or H$_2$O).

We measured the dependence of LIAD yields on the desorbing light wavelength for the Rb-deposited Y surface. Figure \ref{summary} shows a comparatively weak dependence of the LIAD rate on wavelength, compared with quartz surfaces. We also observed an increase in the LIAD rate with Y temperature as shown in Fig.~\ref{Ytemp_dep}, similar to the results obtained for Rb ion-implanted Y foils~\cite{Mar14}. The activation energy was derived as 0.5~eV from the Arrhenius plot in the inset of Fig.~\ref{Ytemp_dep}. These results indicate that the combination of heating and UV light illumination can be used effectively to release alkali-metal atoms, including radioactive Fr atoms, from Y metal.

\begin{figure}
\includegraphics[width=8cm]{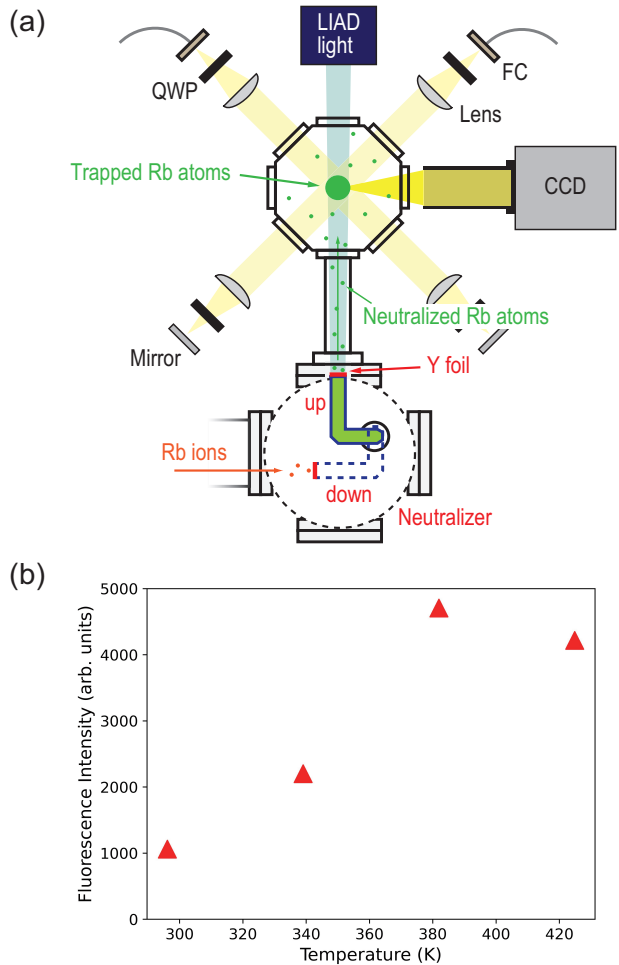}
\caption{(Color online) (a) Experimental setup of the Rb magneto-optical trap (MOT). FC: fiber coupler; QWP: quarter-wave plate; CCD: charge-coupled device camera. (b) Y-temperature dependence of the fluorescence intensity of the Rb MOT loaded by LIAD from the Rb-implanted Y foil using light with a wavelength of 530~nm.}
\label{Rb_MOT}
\end{figure}

We achieved implantation of Rb ions produced in a cyclotron facility, CYRIC at Tohoku University, into a Y foil, then loaded MOT by LIAD from the Y foil at elevated temperatures. The details of the facility and the production beamline have been reported elsewhere~\cite{har16,Dam17}. Figure~\ref{Rb_MOT}(a) shows a schematic diagram of the MOT apparatus. Rb ions were implanted in the Y foil over a 60-s period (``down'' condition). The size of the Y foil was 10 mm $\times$ 15 mm, and the thickness was 0.025 mm. The Y surface was turned to face the glass cell (``up'' condition) and was heated to the desired temperature. The LIAD light then irradiated the Y surface for 60 s. The glass cell was made of quartz and was coated with octadecyltrichlorosilane. The LIAD light wavelength and intensity were 530~nm and 35~mW/cm$^2$, respectively. Rb atoms were released as neutral atoms from the Y surface, because the ionization potential of Rb (4.2~eV) is greater than the work function of Y. The Rb atoms were trapped by MOT. The fluorescence intensity from trapped Rb atoms was detected by a charge-coupled device (CCD) camera. Figure~\ref{Rb_MOT}(b) shows the Y-temperature dependence of the fluorescence intensity of the MOT. The increase in fluorescence intensity due to LIAD from the octadecyltrichlorosilane-coated surface of the upside of the glass cell was estimated from other measurements without the Y foil and then subtracted. The number of atoms in the MOT increased by a factor of 4.5 at 380~K from room temperature, similar to the results shown in Fig.~\ref{Ytemp_dep}; however, the slight decrease at 420~K may be attributed to an increase in the background gas that reduced the trapping time of Rb atoms.

\section{Paraffin}
Coatings are usually used in alkali-metal vapor cells to weaken the atom-surface interactions~\cite{Ste94}. Paraffin, our primary material of interest in this research, consists of long chains of hydrocarbons, rather than oxygen compounds. Its main use in atomic spectroscopy experiments is as an anti-spin-relaxation coating for polarized alkali-metal vapors~\cite{Rob58,Bou66}. We used tetracontane (C$_{40}$H$_{82}$) as a coating material. Tetracontane films were formed by vapor deposition in a manner similar to the approach described in Ref.~\cite{Sek18}. A carbon 1s peak alone was observed by XPS for the formed tetracontane films. This indicates that the coating coverage was very high. Notably, some past XPS studies involving paraffin coatings have demonstrated the XPS peaks of elements originating from the substrates~\cite{Hib13,Kus15}, indicating imperfect or thin coatings.

\begin{figure}
\includegraphics[width=8cm]{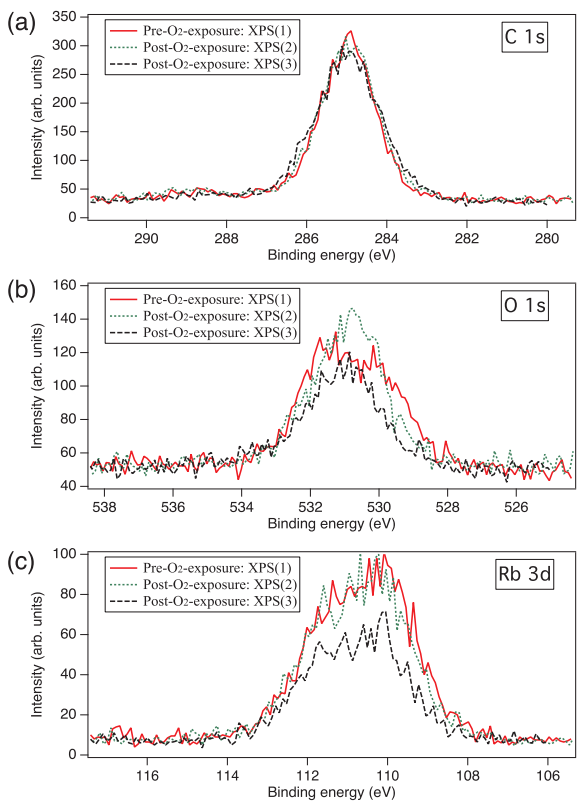}
\caption{(Color online) XPS spectra taken at the times indicated in Fig.~\ref{time-lapse} as XPS(1), (2), and (3) for a tetracontane surface exposed to Rb vapor at 170~K: (a) C 1s peaks, (b) O 1s peaks, and (c) Rb 3d peaks.}
\label{paraffin_xps}
\end{figure}

No Rb atoms detectable by XPS existed after Rb exposure at room temperature. Large amounts of Rb that were observable with the naked eye could be deposited at very low temperatures of approximately 170~K. Similar deposition behavior has been observed for PDMS films. The Rb film formed on tetracontane at low temperatures desorbed with gradual warming to room temperature over a 17-h period, similar to quartz substrates in Sec. III; this desorption was not observable with the naked eye at room temperature. XPS was able to detect the remaining Rb, together with O, as shown in Fig.~\ref{paraffin_xps}. The remaining Rb thus formed compounds with O. Notably, the binding energy was calibrated in the XPS spectra by adjusting the C 1s peak to 285.0~eV. We estimated the concentration ratio of O to Rb to be 1.3 from the peak areas and known photoionization cross-sections, assuming that Rb and O were located at the surface (rather than inside the tetracontane bulk). 

\begin{figure}
\includegraphics[width=8cm]{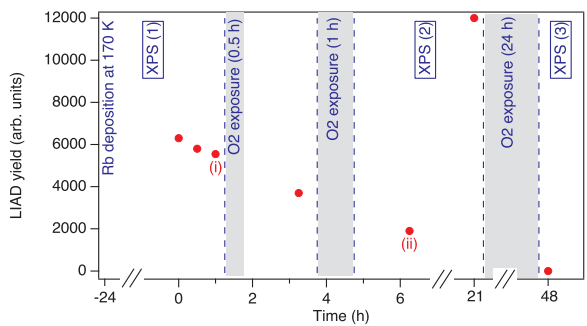}
\caption{(Color online) Time and oxidization dependence of the LIAD yield from a Rb-deposited tetracontane film for LIAD light with a wavelength of 330~nm.}
\label{time-lapse}
\end{figure}

We performed LIAD measurements for the surfaces and confirmed a linear dependence on light power. We then measured the wavelength dependence. To determine the dependence on oxidation, we introduced oxygen gas for a period of 0.5~h or 1~h at a pressure of $1\times10^{-4}$~Pa. The change in the shape of the O 1s peak in XPS spectra, as shown in Fig.~\ref{paraffin_xps}(b), indicates that oxidization proceeded. The concentration ratio of O to Rb also increased to 1.4 for XPS(2). Oxidation likely reduced the LIAD yield, as shown in Fig.~\ref{time-lapse}; we also found a slight change in the wavelength dependence, as shown in Fig.~\ref{summary}.

The status of deposited Rb on paraffin did not exhibit considerable stability. For this series of measurements, the Rb yield was recovered at 21~h, as shown in Fig.~\ref{time-lapse}. The detailed processes leading to this recovery are unclear; however, our results imply that various transportation and chemical processes (e.g., reduction and/or X-ray irradiation) are relevant. The following extended exposure (24~h) to oxygen gas at $1\times10^{-4}$~Pa resulted in no detectable LIAD, whereas XPS confirmed that Rb and oxygen remained on the surface, as shown in Figs.~\ref{paraffin_xps}(b) and (c). The O to Rb ratio increased to 1.5. 

\section{Discussion and conclusions}
The surfaces of many solids used for alkali-metal atomic spectroscopy experiments (e.g., glass and metal) are mainly oxides. Rb deposition onto those surfaces modifies the O 1s XPS peak. This indicates that Rb interacts and forms compounds with the surface O. Polymer surfaces such as PDMS and paraffin are inert to Rb, thus interfering with Rb deposition. However, deposited Rb remains on the surface of O compounds. Using paraffin substrates, we found that the degree of oxidation affected the yield and the wavelength dependence of LIAD; however, the general wavelength-dependence trend did not change. Considering the high reactivity of alkali-metal atoms and prevailing O (including H$_2$O) on the surface and in the gas, our experimental observations imply that some types of compounds of alkali metal and oxygen can be ubiquitous sources for LIAD in various alkali-metal vapor cells, from coated sealed cells to ultrahigh vacuum cells.

It will be informative to investigate the types of alkali-metal oxides that effectively eject alkali-metal atoms on light irradiation, for both practical and fundamental purposes. The ability to control the degree of oxidation on inert surfaces, similar to our approach with paraffin in this study, will be a useful approach. Future studies will provide insights into the preparation of very LIAD-active surfaces and the associated microscopic desorption mechanisms.
 
\begin{acknowledgements}
This work was supported by JSPS KAKENHI Grant Numbers JP23244082, JP26610122, JP26220705, JP16K17676, JP17H02933, JP18K18762, and JP19H05601, and by grants from The Precise Measurement Technology Promotion Foundation and the SEI Group CSR Foundation.
\end{acknowledgements}

\end{document}